# Static flux bias of a flux qubit using persistent current trapping

**Maria Gabriella Castellano[1], Fabio Chiarello[1], Guido Torrioli[1]**
**Pasquale Carelli[2]**

[1] Istituto di Fotonica e Nanotecnologie, CNR, via Cineto Romano 42, I-00156 Roma, Italy
[2] Dipartimento di Ingegneria Elettrica e dell'Informazione, Università dell'Aquila, Monteluco di Roio, I-67040 L'Aquila, Italy

E-mail: mgcastellano@ifn.cnr.it

Qubits based on the magnetic flux degree of freedom require a flux bias, whose stability and precision strongly affect the qubit performance, up to a point of forbidding the qubit operation. Moreover, in the perspective of multiqubit systems, it must be possible to flux-bias each qubit independently, hence avoiding the traditional use of externally generated magnetic fields in favour of on-chip techniques that minimize cross-couplings. The solution discussed in this paper exploits a persistent current, trapped in a superconducting circuit integrated on chip that can be inductively coupled with an individual qubit. The circuit does not make use of resistive elements that can be detrimental for the qubit coherence. The trapping procedure allows to control and change stepwise the amount of stored current; after that, the circuit can be completely disconnected from the external sources. We show in a practical case how this works and how to drive the bias circuit at the required value.

## 1. Introduction

Superconducting qubits based on the magnetic flux degree of freedom are being developed worldwide, with different schemes all based on combinations of Josephson junctions and superconducting loops [1]. These devices behave as an ideal two-state system, described by a symmetric double well potential; symmetry is achieved by a magnetic flux bias of exactly one half of the flux quantum $\Phi_0$=h/2e=2.07 $10^{-15}$ Wb.

A precise and stable flux bias is a key point for the implementation of flux qubits. Precision is needed because even small deviations from $\Phi_0/2$ lead to a substantial asymmetry of the potential that prevents the correct operation as a qubit. Besides, bias fluctuations strongly reduce the qubit coherence time, modulating some of the qubit parameters.

In practical systems, the flux bias is achieved either with a magnetic field externally applied to the qubit [2-4], or by feeding current to a coil inductively coupled to the qubit and integrated on-chip [5-7]. In view of many-qubits applications, the on-chip approach is mandatory: it is scalable, it allows biasing separately each qubit and it makes possible the use of architectures that minimize cross-coupling between lines [7]. However, care must be taken since the feeding circuit can increase the qubit decoherence for several reasons. A first cause is noise coming from the room temperature connection lines; appropriate filtering and shielding can handle this term. Another one is the effect of dissipative parts of the circuits on the qubit. The use of weak coupling between the coils and the qubits [8] is particularly efficient in this respect, but then a large current is required to set the desired flux value. This may produce heating by Joule effect in the low temperature filtering stages, not tolerable in the cryogenic environment used for qubits, in the mK range, where the available refrigerating power is at best tens of microwatts. Using a persistent current, trapped in a superconducting circuit, can solve most of the above problems.

In this paper we propose and demonstrate a circuit formed by a superconducting coil in parallel with an unshunted Josephson junction, which can be used to couple a permanent circulating current to the qubit and approach the required value of flux bias. Briefly, the circuit is a sort of rf-SQUID, whose metastable flux states provide a source of discrete values of circulating current, within a given range. We show how to force the system in a predefined metastable state and so set a particular value of the trapped circulating current, which is then coupled to the qubit: this would be trivial for a shunted rf-SQUID, but it is not so for an unshunted device. After the trapping procedure has been completed, the circuit can be disconnected from the external power supply, with a reduction of the dissipation induced on the qubit; besides, flux quantization in the trapping loop provides bias stability. Fine tuning of the bias flux to $\Phi_0/2$ can be performed as usual with a coupled coil, fed by a small external current through large resistors, without the risk of heating up the system by Joule effect.

## 2. Trapping circuit

The proposed trapping circuit is shown in fig. 1a. The unshunted Josephson junction has a critical current $I_0$ and an effective capacitance $C$; it is in parallel with a superconducting coil with large inductance $L$, used for the current storage. The value of $L$ is such that the reduced inductance $\beta = 2\pi L I_0 / \Phi_0$ is much larger than unity, of the order of hundreds. A smaller coil placed in series with $L$ is used to couple the stored magnetic flux to the qubit. Differently from an ordinary rf-SQUID, the whole circuit is biased by a dc current $I_b$. A similar circuit has been proposed in ref. [9], with the important difference of using shunted junctions for the rf-SQUID: in that case current trapping in the circuit can be controlled very easily, because damping allows the fluxoids to enter the circuit one by one, but shunting resistors do produce decoherence and degrade or forbid qubit performance. It is then necessary to resort to unshunted junctions, which behave differently.

In order to test the behaviour of the unshunted trapping circuit, in this paper we used a slightly different scheme (fig. 1b), where the current stored in the inductance $L$ is read out by a directly coupled dc-SQUID, also unshunted and hence hysteretic.

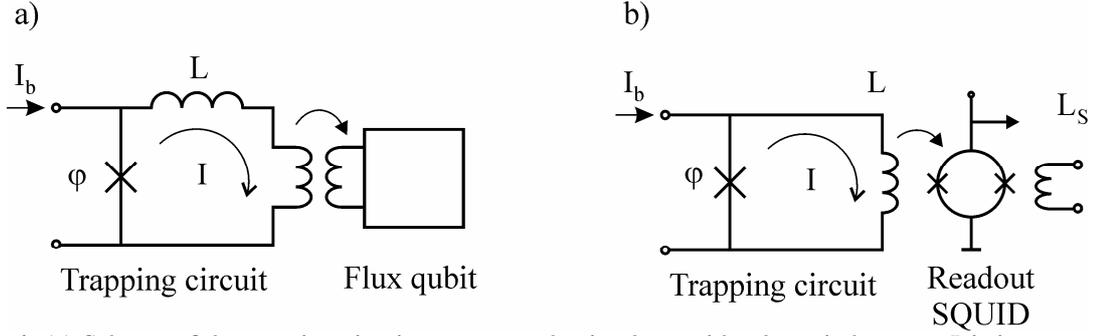

**Figure 1**. (a) Scheme of the trapping circuit: a superconducting loop with a large inductance $L$ is interrupted by a Josephson junction. The circuit is biased by a current $I_b$. A smaller coil, put in series, is used to couple the stored circulating current $I$ to a generic flux qubit. (b) Modified circuit used in this paper: an unshunted readout dc-SQUID is directly coupled to the inductance $L$ to measure the circulating current $I$.

From the Josephson equations, the circuit equation is:

$$I_b = I + I_0 \sin(\varphi) + C\varphi_b \ddot{\varphi} + \frac{\varphi_b}{R} \dot{\varphi} \qquad (1)$$

where $\varphi = LI/\varphi_b$ is the phase difference across the junction, proportional to the current $I$ circulating in the loop, $R$ is the real part of the effective impedance seen by the junction, and $\varphi_b = \Phi_0/(2\pi)$ is the reduced flux quantum. Equation (1) is equivalent to the equation of motion of a particle of mass $C\varphi_b^2$ along the direction $\varphi$ in the potential $U(\varphi)$:

$$U(\varphi) = E_J \left[ -\frac{I_b}{I_0} - \cos(\varphi) + \frac{\varphi^2}{2\beta} \right] \qquad (2)$$

where $E_J = I_0 \varphi_b$, the Josephson energy, sets the energy scale.

For a fixed current bias $I_b$, the potential has the shape of a corrugated parabola, with relative minima separated each other by an energy barrier related to the reduced inductance of the device and to the bias point. Fig. 2 sketches the potential for $I_b=0$ and $\beta=30$; note that this value of $\beta$ is ten times smaller than in our experimental test but it is better suited to illustrate the device behaviour with clarity. The wells represent the available metastable flux states of the system, each corresponding to different values of the trapped circulating current. By changing the bias current the potential rolls on one side, the position of each well is shifted and the barrier between adjacent wells is raised or lowered, depending on their initial position; eventually, a well can become unstable like at point B in fig. 2.

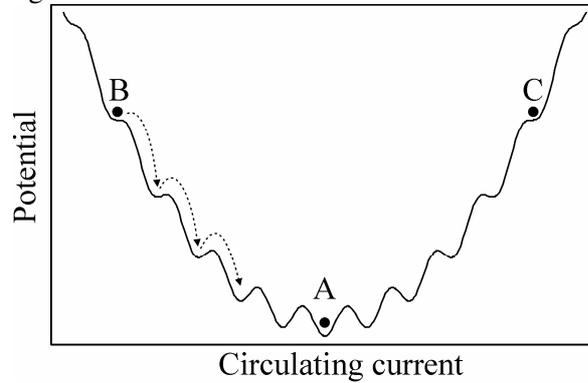

**Figure 2**. Sketch of the potential of the trapping circuit (for fixed current bias $I_b$) versus the circulating current $I$. The points at the relative minima correspond to metastable flux states of the system. The wells at the borders of the figure are shallower and eventually disappear; the corresponding states become unstable (point B).

The position of the critical points of the potential is found by equating to zero the derivative of eq. 2:

$$I_b = I + I_0 \sin\left(\beta \frac{I}{I_0}\right) \qquad (3)$$

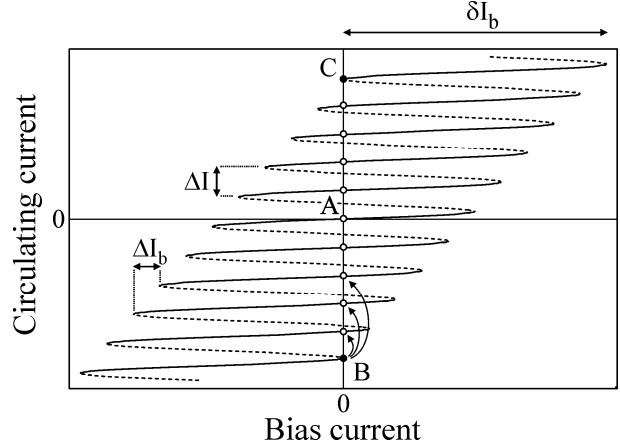

**Figure 3**. Circulating current at the minima (straight lines) and at the maxima of the potential (dashed lines) versus the bias current, for a trapping circuit with β=30. Straight lines represent the possible positions of stable states. The width of each step is $\delta I_b \cong 2I_0$. The two quantities $\Delta I$ and $\Delta I_b$ are equal and given approximately by $\Phi_0/L$. If a system with low damping is in a state that becomes unstable, like in B, it can be retrapped randomly in one of several available states (the empty dots on the vertical line across point B).

Fig. 3 displays the graphical form of eq. 3 with $\beta$=30. Points A, B and C correspond to the same points in the representation on fig. 2. Straight lines ("steps") mark the position of the minima of the potential, at the bottom of the wells. They are almost horizontal, the circulating current $I$ at each of them varying with $I_b$ very weakly as $I \approx 1/(1+\beta)I_b$. The width $\delta I_b$ of the steps sets the range of bias current within which an individual metastable flux state can exist. For a system with $\beta \gg 1$, like in our case, it is $\delta I_b \cong 2I_0$. Dashed lines mark the position of the maxima of the interwell barrier. The corners where maxima and minima coincide correspond to points (like point B in fig. 2) where a minimum disappears and the system can jump to lower energy states. The horizontal and vertical distances between two near steps of fig. 3 are $\Delta I_b = \Delta I \approx \Phi_0/L$ (for $\beta \gg 1$). The total number of wells available at any fixed $I_b$ (or, equivalently, the number of steps that cross the vertical axis at a given $I_b$) is $N \approx \delta I_b / \Delta I_b = \beta/\pi$.

Suppose that the system starts in the deepest well of the potential and that $I_b$ is increased steadily: the potential rolls to the left, the bottom of the initial well is raised and the confining barrier gets smaller. When the local minimum disappears, like at point B of fig. 2, the fictitious particle runs along the potential until it is retrapped in another local minimum. In the language of fig.3 the state moves to the right along a step, towards the edge, where it jumps vertically towards one of the other steps. Actually, the escape from the metastable well is a stochastic process, activated slightly before the disappearance of the potential barrier by thermal fluctuations or quantum tunnelling [10, 11]. This leads to a "gray zone", a range of values of $I_b$, close to the edge, where transition can occur; the amplitude of the gray zone is related to the effective temperature or to quantum effects. At the low temperature used in this work, the effect is quite small and it will be neglected in the general explanation of the circuit operation.

If the system is sufficiently damped (namely if the McCumber parameter $\beta_c = I_0 R^2 C / \varphi_b \leq 1$, as it may happen if the Josephson junction is externally shunted with a low value resistor), the escape from the metastable state is followed by an immediate retrapping in the closest potential well. In other words, there are jumps only between adjacent steps of fig. 3 and fluxoids enter the circuit one by one [9]. If instead the damping is low ($\beta_c \gg 1$), like for our unshunted circuit, the retrapping may occur randomly even in a metastable state far from the initial one, depending on the balance between gained kinetic energy and dissipation. For instance, starting from point B in fig. 2 the system can end up in any of the potential wells between B and C; equivalently, in fig. 3 any of the empty circles represent possible arrival states and multiple jumps are observed (curved arrows). It should be noted that the arrival point is stable against thermal or quantum fluctuations only if it does not lie in the gray zone, so that the available range for current storage is slightly reduced with respect to the ideal case that doesn't consider stochastic effects.

In summary, for any fixed value of the bias current, a current within the range $\pm I_0$ can be stored in the circuit, with increments of $\sim\Phi_0/L$. Each value of the trapped current corresponds to a different metastable state (the wells in fig. 2 or equivalently the steps in fig.3) and which one is chosen depends on the previous history of the system. Metastable states are travelled one by one only if the damping of the system is sufficiently high; when there is negligible damping, the arrival state (and hence the current trapped in the system) is randomly chosen in a range of possibilities. However, by using the trapping procedure discussed in the next paragraph, this behaviour can be avoided and a specific arrival state can be selected, without resorting to increased damping that would badly affect the qubit performance.

### 3. Trapping procedure

In order to bias the qubit at the desired flux point, the trapping circuit must be driven at the correct state, choosing one specific metastable flux state among the many compatible with that particular value of the bias current. The selection of one particular well as working point in the multiwell potential of a Josephson device is a topic that has been discussed in the literature, especially for defining the initial state of a Josephson qubit. Different strategies have been envisaged for the two-dimensional potential of a dc-SQUID. In the "forced-retrapping" scheme of ref. [12], an oscillating bias current is applied to a hysteretic dc-SQUID to de-trap the SQUID out of all but the deepest potential well. Very recently [13], a "flux-shaking" technique, acting on the flux instead of the bias current, has been used to select any of the allowed flux states of a dc-SQUID. Our trapping procedure is an extension of these schemes, which is described and applied here for the first time to the one-dimensional potential of an rf-SQUID.

We start with zero $I_b$ and we proceed to store a current $I_x$ (included within the allowed range $\pm I_0$) in the trapping circuit. At $I_b = 0$ the corresponding metastable state will be a relative minimum of the potential. In order to force the system in this well, a dc bias current is applied and the potential is tilted until the selected well becomes the absolute minimum. The potential is such that this happens for $I_b \sim I_x$. Adding a sinusoidal signal to the bias current, the potential is shaken so that any other state than the absolute minimum becomes unstable at some point of the shaking. The sweeping signal has a frequency ranging from hundreds Hertz to kHz and is left acting for several cycles, during which the system has the time to relax in the deepest well and stay there. The optimal peak-to-peak amplitude of the sweeping signal is equal to the quantity $\delta I_b \cong 2I_0$, so that all the existing metastable states are involved in the shaking procedure. If the sweeping amplitude is larger than the optimum, at some point of each cycle even the initial, deeper well becomes unstable. Smaller amplitude, on the other hand, permits to a larger number of states to remain stable against the sweep, with the result that the system can land and remain in more than a single well, at random. At the end of the sweeping procedure, the dc offset is removed, $I_b$ is again set to zero and the circuit can be disconnected from the external source. Consequently the potential returns to its initial shape, but the system remains trapped in the selected well, with the desired circulating current stored in the loop.

### 4. Experimental results

The device we tested was fabricated by PTB, Braunschweig, Germany, using Nb trilayer technology with a critical current density of 100 A/cm$^2$; the junctions were defined by anodization and dry etching, with SiO$_2$ insulator. The circuit for flux trapping consists of a multiturn coil with total inductance $L=1.5$ nH, in parallel with an unshunted Josephson junction of 10 µm nominal diameter, whose measured critical current resulted to be $I_0=100$ µA. The reduced inductance for the system should then be $\beta\sim470$, the number of available metastable states $\beta/\pi\sim150$ and the difference in the current stored in two nearby states $\Phi_0/L\sim1.4$ µA. The system is read out by an unshunted two-hole dc-SQUID, with gradiometric structure, inductance of a few pH and critical current of 22 µA. In this test, one hole of the dc-SQUID is directly coupled to the inductance $L$, while the other is coupled to a secondary coil, identical to that of the flux trapping circuit, used for the flux bias of the dc SQUID readout. The coupling between each of the coils and the dc SQUID is $M=120$ pH, corresponding to a current $\Phi_0/M=17.6$ µA needed to produce one flux quantum in the readout SQUID. Such a large coupling, chosen here to make the measurements easier, is not necessary in a practical use.

A photograph of the device is shown in fig. 4a. It must be noted that in the view of an integration with a real qubit, the device size can be made much smaller while keeping the same parameter values, as shown in the layout in fig. 4b. In this layout, the feature size is compatible with typical foundry rules; the linewidth is 2 µm, the spacing is 2.5 µm. The inductance $L$ is built with gradiometer structure; this allows to reduce (about a factor 100) the pickup of environmental magnetic fluctuations that could affect the flux stability. The device has to be connected to a properly designed coil that couples flux to a specific qubit. With our present design, where the current flowing in the circuit is limited to about $\pm100$ µA, a mutual inductance of 20 pH is required to couple $\pm1\Phi_0$ in the qubit. This figure is easily achievable if the qubit loop has a large size (about 100 µm), like in [5], [7], [11]: in this case one can use an overlapping coil or a straight line close enough (a few micrometers) to the loop side. In the case of much smaller qubits, like in the persistent current (PC) qubits, where the typical loop size is 10 µm, it is necessary to enhance the coupling resorting to a coil surrounding the qubit; this solution is commonly adopted in PC qubits for the readout dc-SQUID or for the readout tank circuit. As an alternative, one can use a larger junction or a parallel array of several junctions in the trapping circuit, in order to achieve a larger stored current.

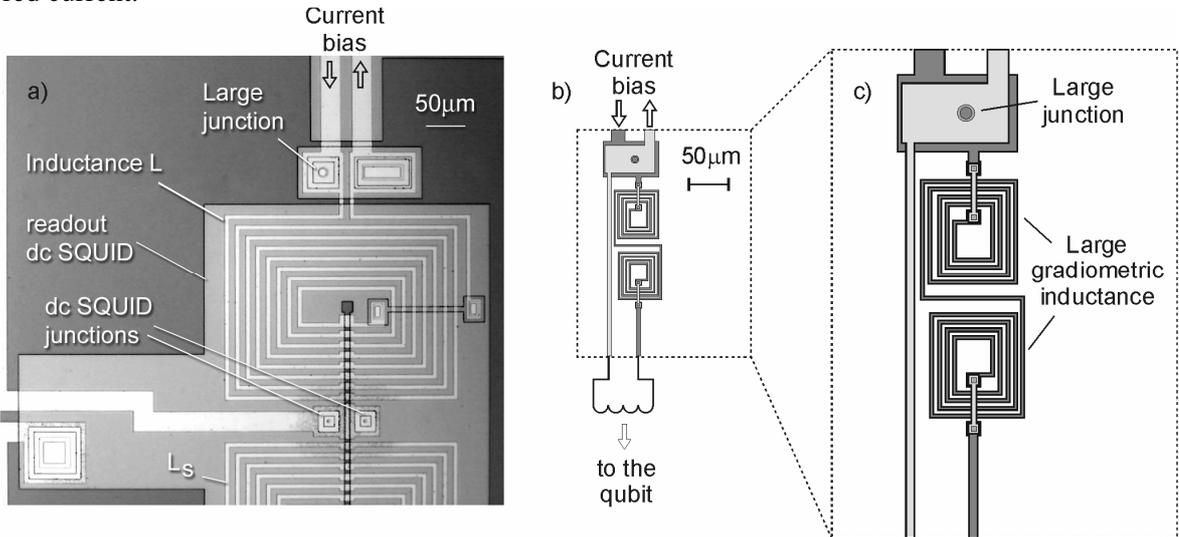

**Figure 4**. (a) Photograph of the circuit measured in this paper, corresponding to the scheme of fig. 1b. (b) Possible layout for a trapping circuit of reduced size, to be connected to flux qubits. (c) Enlargement of the core of fig. 4b.

The measurements have been performed at 380 mK, in a $^3$He refrigerator (Heliox by Oxford Instruments), with the chip enclosed in a well shielded copper case, with filters to protect it against external electromagnetic noise.

After applying the trapping procedure, the device response is tested by monitoring the modulation curve of the readout dc-SQUID (critical current versus magnetic flux) and recording the flux displacement due to the circulating current stored in the trapping circuit. Since the dc-SQUID is hysteretic, the critical current is measured by sweeping the dc-SQUID with a current ramp, typically at a frequency of 5-10 kHz. When the bias reaches the critical current value, the dc SQUID undergoes a transition from the superconducting to the dissipative state and the voltage appearing across the SQUID triggers the acquisition of the instantaneous value of the bias current. The process is repeated for several cycles (N=10÷1000) and a mean value of the switching current is recorded. In order to trace the modulation curve, the procedure is repeated for different values of the magnetic flux coupled to the dc-SQUID by its ancillary coil. The flux coupled by the trapping circuit is detected by measuring the horizontal shift produced on the whole modulation curve.

In the readout dc-SQUID, the process of escape from the zero-voltage state is stochastic and depends on noise, thermal fluctuations and quantum tunnelling; hence, the instantaneous values of the dc-SQUID critical current is spread over a range of values, upper-limited by the zero-fluctuation values. As a consequence, the modulation curve has a finite width that limits the sensitivity with which the horizontal shift of the whole curve can be identified and measured. By converting the measured critical current fluctuation to magnetic flux through the dc-SQUID responsivity, we find a sensitivity $\Delta\Phi \sim 3.5$ m$\Phi_0$. This quantity is further reduced by a factor $N^{1/2}$ because the modulation curve is traced not with the istantaneous switching current but with the average over N cycles, so that in our case $\Delta\Phi \leq 1$m$\Phi_0$.

As described before, current is stored in the trapping loop by applying a dc current bias with a superposed sine waveform (0.1-1 kHz), while keeping the readout SQUID off. The optimal peak-to-peak amplitude, found experimentally, is 200 µA and confirms the value of the junction critical current, $I_0$=100 µA. After about 200 oscillations the system relaxes on the desired metastable well, the one corresponding to the desired value of circulating current. The bias current is then switched off and leaves a persistent current stored in the trapping circuit, about the value of the dc offset that was applied previously. The readout SQUID, which was off during the trapping procedure, is turned on and the flux shift of its modulation curve is measured and plotted against the stored current. The procedure is repeated for several values of the trapped current. To give an idea of the reproducibility of the process, it was possible to turn off the system, make a thermal cycle at 4.2 K, wait about one day with the sample at 4.2 K, cool down again, repeat the trapping procedure with the same bias values and find the system in the same state as the day before: that is, the current-flux characteristics of the readout dc-SQUID were not shifted with respect to the previous day, within the experimental sensitivity.

Fig. 5 shows the stored current as a function of the offset value $I_b$ that was applied in the trapping procedure. The stored current is measured through the flux displacement that it produces in the readout dc-SQUID (right scale of fig. 5) and can be easily converted in Amps (left scale of fig. 5) by using the measured mutual inductance $M$. The graph shows an apparent linear increase for a range of $I_b$ equal to 176 µA, about 10% lower than the maximum allowed $2I_0$= 200 µA (as expected because of instabilities due to the gray zone). This is also the corresponding range of the stored current. As shown in the enlargement, the apparent steadily increase is actually a sequence of equal steps, showing that fluxoids enter the system one by one (i.e., metastable states are travelled one by one): by increasing the bias offset, the system remains in the same metastable state for a small range of values until it switches to the next one. For values of $I_b$ external to the allowed interval $\pm I_0$, the system is found each time in a different metastable state, at random.

The measured number of steps in the allowed region is about 140, close to the prediction. The vertical spacing between two adjacent steps corresponds to a shift of 75 m$\Phi_0$ in the readout dc-SQUID. Divided by the mutual inductance $M$, this gives a value of 1.3 µA for the discrete increment of the corresponding trapped current, close the expected 1.4 µA. The statistical analysis of the measured vertical spacings shows that 130 of them have the same value within 1 m$\Phi_0$, while the remaining ones have a double value. The reason for these double jumps (shown in the right enlargement of fig. 5) seems to be related to the interaction with the readout dc-SQUID; as a fact, the periodicity of the double jumps is the same as the flux periodicity of the readout dc-SQUID.

Finally, we checked experimentally the dependence of the curve of fig. 5 on the sweeping amplitude used in the trapping procedure and set the optimal amplitude by optimizing the output curve: only the correct amplitude gives a uniform sequence of equal steps, with minimal number of multiple jumps.

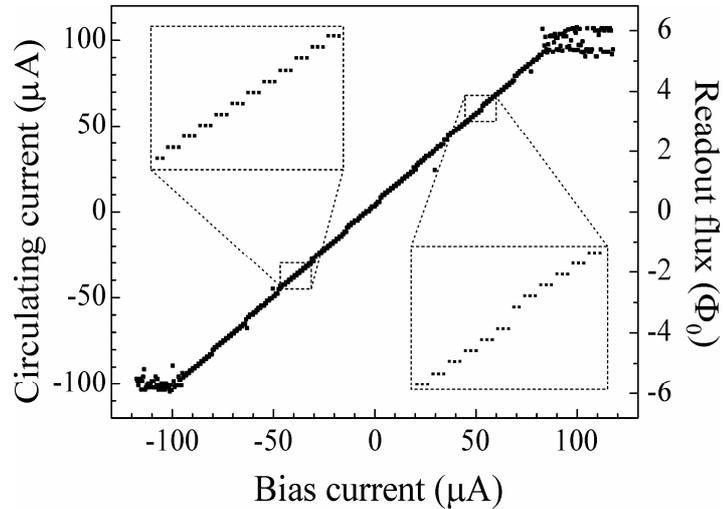

**Figure 5.** Current stored in the trapping circuit for different dc values of the bias current $I_b$ applied in the trapping procedure. The y-axis is measured with two scales: flux displacement in the readout SQUID (in $\Phi_0$) and circulating current (in μA). The curve is a sequence of steps (see enlargements), each corresponding to a fluxoid entering the trapping circuit, with a vertical increment of 1.3 μA that represent the quantization step for the trapped current. The right enlargement shows the presence of double jumps, one for each flux quantum flowing in the readout dc-SQUID; this effect is probably related to the back-action of the dc-SQUID on the trapping circuit.

### 5. Discussion and conclusion

We have proposed and demonstrated a method to flux-bias flux qubits with a permanent current stored in superconducting circuit (basically an unshunted rf-SQUID), inductively coupled to the qubit. The goal is to approach as close as possible the desired flux bias value with a system that provides stable values, does not add dissipation and does not heat up the system; all these requirements are satisfied by a supercurrent trapped as discussed.

Decoherence induced by circuits close to the qubit is a key point in the project of a quantum computing system. The general strategy to minimize decoherence for flux qubits [14] is that the lines and circuits close to it must have large real impedance and very weak inductive coupling. The required values are depending on the particular type of qubit but a typical order of magnitude is 100 Ω for the resistors and a few picoHenrys for the mutual inductance. The use of our scheme to flux bias the qubit allows to relax some of these conditions. A first advantage is that after the trapping procedure has been completed, the trapping circuit can be disconnected from the external bias circuitry. Moreover, any noise contribution coming from the remaining lines of the trapping circuit is transferred to the qubit through the circulating current with a reduction of $1/(1+\beta)$, due to the shape of the transfer function (fig. 3). In our case this leads to a reduction by a factor of about 500 of the effective coupling for noise, while still allowing an efficient coupling for static flux bias of the qubit. If most of the flux bias is supplied by the trapped current, the additional bias for fine tuning must be only a small correction, within one or two quantization steps of the trapped circulating current. For instance, with a quantization step of 1.3 μA and a mutual inductance of 20 pH like in our case, the flux for fine tuning must be at most 30 m$\Phi_0$, instead of the 500 m$\Phi_0$ needed in the absence of a trapping circuit. This allows the use of smaller inductive coupling for these additional lines, again relaxing the filtering requirements in this case by a factor 10.

In conclusion, in this paper we reported measurements at 380 mK on a prototype where the trapping circuit was read out by a hysteretic dc-SQUID. The tests have shown that, with our technique, a current as large as $\pm I_0 \sim \pm 100$ μA can be trapped in the loop, and that the stored values can be varied at will with a quantization step of 1.3 μA. These figures can be easily changed by choosing the value of the circuit parameters, to tailor the technique to the specific qubit requirements.

We gratefully acknowledge A. B. Zorin, M. Khabipov and D. Balashov at PTB, Braunschweig, for fabricating the chips. We thank R. Leoni and C. Cosmelli for discussions and suggestions. This work is supported by the European Community Project RSFQubit (FP6-3748).